\documentclass[epj]{webofc}
\usepackage[utf8]{inputenc}
\usepackage[varg]{txfonts}   
\usepackage{booktabs}
\usepackage{xcolor}
\definecolor{darkred}{rgb}{0.4,0.0,0.0}
\definecolor{darkgreen}{rgb}{0.0,0.4,0.0}
\definecolor{darkblue}{rgb}{0.0,0.0,0.4}
\usepackage[bookmarks,linktocpage,colorlinks,
    linkcolor = darkred,
    urlcolor  = darkblue,
    citecolor = darkgreen]{hyperref}
\usepackage{amsmath}
\usepackage{slashed}
\wocname{EPJ Web of Conferences}
\woctitle{Lattice2017}
\newcommand{\eq}[1]{eq.~(\ref{#1})}

\newcommand{\ri}{{ \mathrm{RI} }}
\newcommand{\tree}{{ \mathrm{tree} }}
\newcommand{\lat}{ {\mathrm{lat}} }
\newcommand{\ms}{{\overline{\mathrm{MS}}}}

\newcommand{\fm}{\mathrm{fm}}
\newcommand{\GeV}{\mathrm{GeV}}
\newcommand{\qcd}{ {\mathrm{QCD}} }

\newcommand{\GF}{G_\mathrm F}

\newcommand{\mw}{ {m_\mathrm W} }

\newcommand{\VmA}{{\mathrm V - \mathrm A}}
\newcommand{\Sp}[1]{S(#1)}

\newcommand{\ZV}{Z_\mathrm V}

\begin{document}
%
\selectlanguage{english}
\title{%
Weak hamiltonian Wilson Coefficients from Lattice QCD
}
\author{%
\firstname{Mattia}  \lastname{Bruno}\inst{1}\fnsep\thanks{Speaker, \email{mbruno@bnl.gov}}
}
\institute{%
Physics Department, Brookhaven National Laboratory, Upton, NY 11973, USA
}
\abstract{%
In this work we present a calculation of the Wilson Coefficients 
$C_1$ and $C_2$ of the Effective Weak Hamiltonian to all-orders in 
$\alpha_s$, using lattice simulations. 
Given the current availability of lattice spacings 
we restrict our calculation to unphysically light 
$W$ bosons around 2 GeV and we study 
the systematic uncertainties of the two Wilson Coefficients.
}
\maketitle

\section{Introduction}

Weak decays of hadrons have a rich phenomenology and their study is important
to continue to test the validity of the Standard Model.
In these processes two fundamental scales are involved: the typical size of 
QCD bound states, such as the light mesons, of $O(\Lambda_\mathrm{QCD})$ 
and the mass of the weak bosons that mediate the decays.
Therefore in the presence of a natural scale separation,
the effective field theory environment is the ideal ground to provide
a theoretical prediction for these decays. 
For simplicity let us concentrate on a transition of the form $c \to s u \bar d$.
By integrating out the
heavy degrees of freedom, in particular the weak bosons and the heavy quarks, one obtains a new
effective theory with four-quark interactions that require a proper renormalization
and whose strength is given by the new coupling constants of the effective theory:
the so-called Wilson Coefficients. 
The corresponding effective hamiltonian reads as follows
\begin{equation}
\mathcal{H}_\mathrm{eff} = V_\mathrm{cs} V_\mathrm{ud} \frac{\GF}{\sqrt{2}} \sum_{i=1,2} C_i Q_i \,,
\label{eq:heff}
\end{equation}
with the dimensionful Fermi constant $\GF$ and the two current-current operators 
\begin{equation}
Q_1 =  (\bar s_i c_j)_{\VmA} (\bar u_j d_i)_{\VmA} \,, \quad 
Q_2 =  (\bar s_i c_i)_{\VmA} (\bar u_j d_j)_{\VmA} \,.
\end{equation}
The flavor structure of the decay prevents the appearance of penguin diagrams in the full 
theory and correspondingly of disconnected diagrams in the EFT.
In the rest of this work we will focus on this type of transitions only.
When the operators are evaluated within certain external states and renormalized at a scale 
$\mu$, the scale separation emerges naturally: 
the Wilson Coefficients $C_1$ and $C_2$ contain the information related to the short-distance
part of the process with energies above $\mu$ and the renormalized matrix elements 
describe the long-distance contributions with energies below $\mu$. 
For more details on the weak effective hamiltonian see the comprehensive review in 
Ref.~\cite{buras95}.

In the last decade, the Lattice QCD Community 
made significant progresses in the calculation of the matrix
elements with mesonic states at physical kinematics for several processes: the RBC 
collaboration recently completed the first full calculation of 
the decays of a kaon into two pions in the two isospin channels~\cite{Bai:2015nea, Blum:2012uk, Blum:2011ng,CKelly}. Among the
usual uncertainties associated with these type of calculations, a significant portion of the error
still comes from the Wilson Coefficients, whose perturbative knowledge is currently limited 
to next-to-leading order~\cite{buras95} and NNLO for some of them~\cite{Gorbahn:2004my}. 
In this work we try to define a strategy to compute them to all-orders in the strong coupling
constant through lattice simulations. In the next section we review the main obstacles
and difficulties of this calculation on the lattice, while in the third section
we outline the
computational method. In section 4 we present preliminary results 
before concluding.

\section{A non-perturbative calculation of the Wilson Coefficients}

The main goal of the present study is to define a strategy to compute the 
initial conditions of the Wilson Coefficients using lattice QCD, to obtain a 
result to all-orders in the strong coupling constant.
From the matrix elements computed by the RBC/UKQCD collaboration for the 
$I=0$ channel of the $K \to \pi \pi$ decay~\cite{Bai:2015nea} we have examined
the systematic error introduced from the Wilson Coefficients alone, which amounts to
approximately 12\%.
Starting from the initial conditions at the $W$ pole, the Wilson Coefficients are 
evolved to lower scales with the step-scaling matrices and matched at various
quark thresholds between theories with different numbers of active flavors.
By looking at the difference of the LO and NLO approximations of the initial conditions
we have estimated an effect on the imaginary part of the amplitude 
of appriximately 3\% for $C_1$ and $C_2$ and up to 6\% for the entire basis.
For these reasons we have initiated an effort to improve the determination
of the initial conditions of the Wilson Coefficients. 
In Ref.~\cite{Dawson:1997ic} a similar strategy
to define a non-perturbative weak hamiltonian was proposed.

Now let us examine a few caveats of the results presented below. 
For simplicity we restrict ourselves to the current-current operators in the EFT, as already
described in the introduction, thus avoiding the penguin and disconnected diagrams which are
more difficult to treat on the lattice. Therefore only the real parts of the $K \to \pi \pi$ 
amplitudes would benefit from this study.
Moreover due to the current limitations in the availability of fine lattice spacings, for large
volume simulations, we have studied an unphysical scenario with light $W$ bosons of masses
around $2~\GeV$. We have adopted Regularization Independent renormalization conditions in
momentum space (RI/(S)MOM) and we have studied the dependence on the volume of the lattice and
on the quark mass of our Wilson Coefficients.
This is important to ensure that the EFT with the dimension-6 operators only, reported in \eq{eq:heff},
provides a reliable description of the physical process. Higher dimensional operators
are expected to be of $O(\mu_0^k/ \mw^k)$, with $k>0$ and $\mu_0$ a generic infrared scale 
of the system, such as the typical momentum of the process, the quark mass, 
the finite box size or $\Lambda_\qcd$.

\section{The main strategy}

In this section we present our non-perturbative calculation of the initial conditions
of the Wilson Coefficients. Our approach is based on the RI/(S)MOM 
renormalization scheme~\cite{Martinelli:1994ty,Aoki:2007xm,Sturm:2009kb}.
The building blocks of the amputated Green's functions are the quark 
propagators in momentum space, which we obtain by inverting the Dirac operator $D(x,y)$ 
on plane waves with momentum $p$ 
\begin{equation}
G(x,p) = \sum_y D^{-1}(x,y) e^{ipy} \,,
\label{eq:Gxp}
\end{equation}
in Landau gauge.
By multiplying $G(x,p)$ with the appropriate phase factor
we define
\begin{equation}
\tilde G(x,p) = e^{-ipx} G(x,p) \,, \quad \Sp{p} = \frac{1}{V} \sum_x \langle e^{-ipx} G(x,p) \rangle \,,
\label{eq:props}
\end{equation}
that we use to construct the amplitudes in the EFT
involving the two four-quark operators: we provide below 
the full expression for $Q_2$, where we omit the indices within the square brackets
(greek and roman letters label spin and color indices respectively)
\begin{equation}
[\Gamma(Q_2) ]^{\alpha \beta \gamma \delta}_{a b c d} (p_1,p_2,p_3,p_4) = \sum_{\mu \,, x} 
 \langle \big[ \gamma_5 \tilde G(x,-p_2)^\dagger \gamma_5 \gamma_\mu^L \tilde G(x,p_1) \big]^{\alpha \beta}_{a b} 
 \big[ \gamma_5 \tilde G(x,-p_4)^\dagger \gamma_5 \gamma_\mu^L \tilde G(x,p_3) \big]^{\gamma \delta}_{c d} \rangle \,.
\label{eq:gamma_Q2}
\end{equation}

On the full theory side only a single diagram is needed. To compute the matching, we use
the same quark propagators of \eq{eq:gamma_Q2}.
The amplitude that we evaluate in the full theory involves the $W$ boson
propagator at tree level in unitary gauge (and Euclidean space-time)
\begin{equation}
W_{\mu \nu}( \hat p) = \frac{1}{ (a\hat p)^2 + (a\mw)^2} 
\Big( \delta_{\mu\nu} - \frac{\hat p_\mu \hat p_\nu}{\mw^2} \Big) \,,
\quad
a \hat p_\mu = 2 \sin ( a p_\mu /2 ) \,,
\label{eq:wprop}
\end{equation}
which we fourier-transform to coordinate space and use inside the following equation
\begin{equation}
\begin{split}
[\Gamma_\mathrm{SM} ]^{\alpha \beta \gamma \delta}_{a b c d} (p_1,p_2,p_3,p_4) = & \sum_{\mu \nu} \sum_{x,y} 
 \langle \big[ \gamma_5 \tilde G(x,-p_2)^\dagger \gamma_5 \gamma_\mu^L \tilde G(x,p_1) \big]^{\alpha \beta}_{a b} \ W_{\mu \nu}(x,y) \\ 
 & \times \big[ \gamma_5 \tilde G(y,-p_4)^\dagger \gamma_5 \gamma_\nu^L \tilde G(y,p_3) \big]^{\gamma \delta}_{c d} \rangle \,.
\label{eq:Wamp}
\end{split}
\end{equation}

To amputate the above Green's functions the inverse of the expectation value
of the appropriate quark propagators are multiplied accordingly.
We define the amputated Green functions in the EFT as
$\Lambda(Q_i)$ and in the full theory as $\Lambda_\mathrm{SM}$.
They can be projected 
onto definite spin-color states using
\begin{equation}
P_1 = \delta_{bc} \delta_{da} \, \Gamma_1 \otimes \Gamma_2 \,, \quad
P_2 = \delta_{ba} \delta_{dc} \, \Gamma_1 \otimes \Gamma_2 \,,
\label{eq:proj} 
\end{equation}
whose color structure resembles the one in the operators. 
More precisely by defining
the action of a projector onto an amputated amplitude, we construct the matrix $M$ as
\begin{equation}
M_{ij} = P_j[ \Lambda(Q_i) ] \,, \quad P_j[ \Lambda(Q_i) ] \equiv \mathrm{Tr} (P_j \Lambda(Q_i)) \,,
\end{equation}
where the trace runs over spin and color indices.
With the same projectors we define the vector 
$W_i \equiv P_i[ \Lambda_\mathrm{SM} ]$.
Note that we have the freedom to choose an arbitrary combination 
of spin matrices inside \eq{eq:proj} as long as it provides an invertible matrix once applied to the
Green's functions. We exploit this freedom and we examine two possibilities: one given
by the combination $\mathrm{VV}+\mathrm{AA}$, thus preserving parity, and the second
given by the parity odd combination $\mathrm{VA}+\mathrm{AV}$.
In our notation $\mathrm{VA}$, for example, represents the following tensor product
$\sum_\mu \gamma_\mu \otimes \gamma_\mu \gamma_5$, which also defines the so-called $\gamma$ 
projectors. Alternatively, replacing $\gamma_\mu$ with $(\slashed p p_\mu)/p^2$
and $\gamma_\mu \gamma_5$ with $(\slashed p \gamma_5 p_\mu)/p^2$ defines the 
$\slashed p$ scheme~\cite{Lehner:2011fz}.

At this point we can impose the usual RI conditions on the 
amputated and projected amplitudes~\cite{Martinelli:1994ty}
\begin{equation}
\lim_{m_q \to 0} \
\frac{Z_{ij}^\ri}{(Z_q^\ri)^2} \,
M_{jk}^\lat|_{\mu^2=p^2} = M_{ik}^\ri |_{\mu^2=p^2} \equiv M_{ik}^\tree \,,
\label{eq:rimom}
\end{equation}
with $Z_q^\ri$ being the wave function renormalization.
The remaining degree of freedom that we can explore is the momentum
configuration used in the definition of the amplitudes, namely the
four momenta used in the four external off-shell quark legs.
In the present work we study both the exceptional, with $p_1=p_3=p$ and $p_2=p_4=-p$,
and non-exceptional kinematics, given by
\begin{equation}
ap_1 = (x,-x,0,0) \quad ap_2 = (0,0,-x,x) \quad
ap_3 = (-x,0,x,0) \quad ap_4 = (0,x,0,-x) \,,
\label{eq:nonex}
\end{equation}
leading to momentum conservation of the amplitudes ($x$ is a generic dimensionless parameter).
The full theory does not require additional renormalization factors, such as
the four-quark matrix $Z$, besides $Z_q^\ri$.
Nonetheless, the usage of vector and axial local currents 
with the lattice regulator demands a finite normalization factor $Z_V$ (or $Z_A$)
and we take that into account.\\

Now we have all the ingredients to move forward and 
perform the matching between the full theory and the effective hamiltonian: 
we simply need to equate the renormalized effective hamiltonian and the full theory, 
with the appropriate normalization factors and couplings
\begin{equation}
\frac{\GF}{\sqrt{2}} \ C_i^{\ri}(\mu) M_{ij}^{\ri}(\mu) = W_j^{\ri}(\mu) =
\frac{g_2^2}{8} \frac{Z_V^2}{(Z_q^\ri)^2} W_j^\lat \,.
\label{eq:matching}
\end{equation}
In \eq{eq:matching} $\GF$ represents the dimensionful Fermi constant and $g_2$ the
bare weak coupling. The two simplify leaving only a factor $\mw^2$ on the r.h.s.
of \eq{eq:matching}.
Expanding $M^\ri$ according to \eq{eq:rimom} 
and solving this equation for the Wilson Coefficients leads to 
\begin{equation}
C_i^\lat \equiv \ \mw^2 \Big( W_j^{\lat} [M^{\lat}]^{-1}_{ji} \Big) \,, 
\qquad C_i^{\ri}(\mu) = \ C_j^\lat  \
\Big( [Z^{\ri}(\mu)]^{-1}_{ji} Z_V^2 \Big) \,.
\label{eq:wcoeff}
\end{equation}
In \eq{eq:wcoeff} we have explicitly separated the calculation of the bare lattice
Wilson Coefficients $C_i^\lat$ from their renormalization: this allows us to carefully study
the matching on the lattice as a function of $p^2/\mw^2$ ($p$ being the momentum of the 
external quark states);
once this is achieved, we perform a separate calculation at higher scales,
with $p$ of $O(\mw)$, from which we estimate the renormalization matrix $Z^\ri$ together
with the finite normalization factor $\ZV$.

In a periodic box, momenta are quantized in units of $2\pi / L$. 
To overcome this restriction, especially
at small momenta, 
we have adopted
twisted boundary conditions, introduced in Ref.~\cite{deDivitiis:2004kq} and successfully applied 
in Ref.~\cite{Arthur:2010ht} for quantities similar to the ones studied in this work.
By measuring
our amplitudes always in a given irreducible representation of $H_4$, we suppress 
hypercubic breaking effects.

\section{Preliminary results}

The results presented below are based on three ensembles
generated with the Iwasaki gauge action and 2+1 Shamir Domain Wall
fermions in the sea\footnote{The extent of the fifth dimension is 16 sites.}.
Two ensembles share the same gauge coupling $\beta=2.13$ and 
bare quark masses 0.01 and 0.04 and differ only in the physical volume: we label 16I 
the $16^3 \times 32$ lattice and 24I the $24^3 \times 62$ lattice.
Their lattice spacing in physical units is approximately 1.8 $\GeV^{-1}$.
The third ensemble with volume $32^3 \times 64$, which we denote with 32I, 
is used to take the continuum limit, since the gauge coupling $\beta=2.25$
corresponds to $a^{-1} \approx 2.4 ~\GeV$.
More details on other algorithmic and physical parameters can be found in 
Refs.~\cite{Aoki:2010dy,Allton:2007hx}.
On the 16I and 24I we have performed measurements up to $0.8~\GeV$, 
whereas on the 32I up to $0.4 ~\GeV$.
On each configuration 4 different inversions are required (one per momentum) 
to construct the non-exceptional kinematics. We have used $O(30)$ configurations
per ensemble, separated by at least 100 Molecular Dynamics Units.

\subsection{Higher order operators}

\begin{figure}[ht]
\includegraphics[width=.49\textwidth]{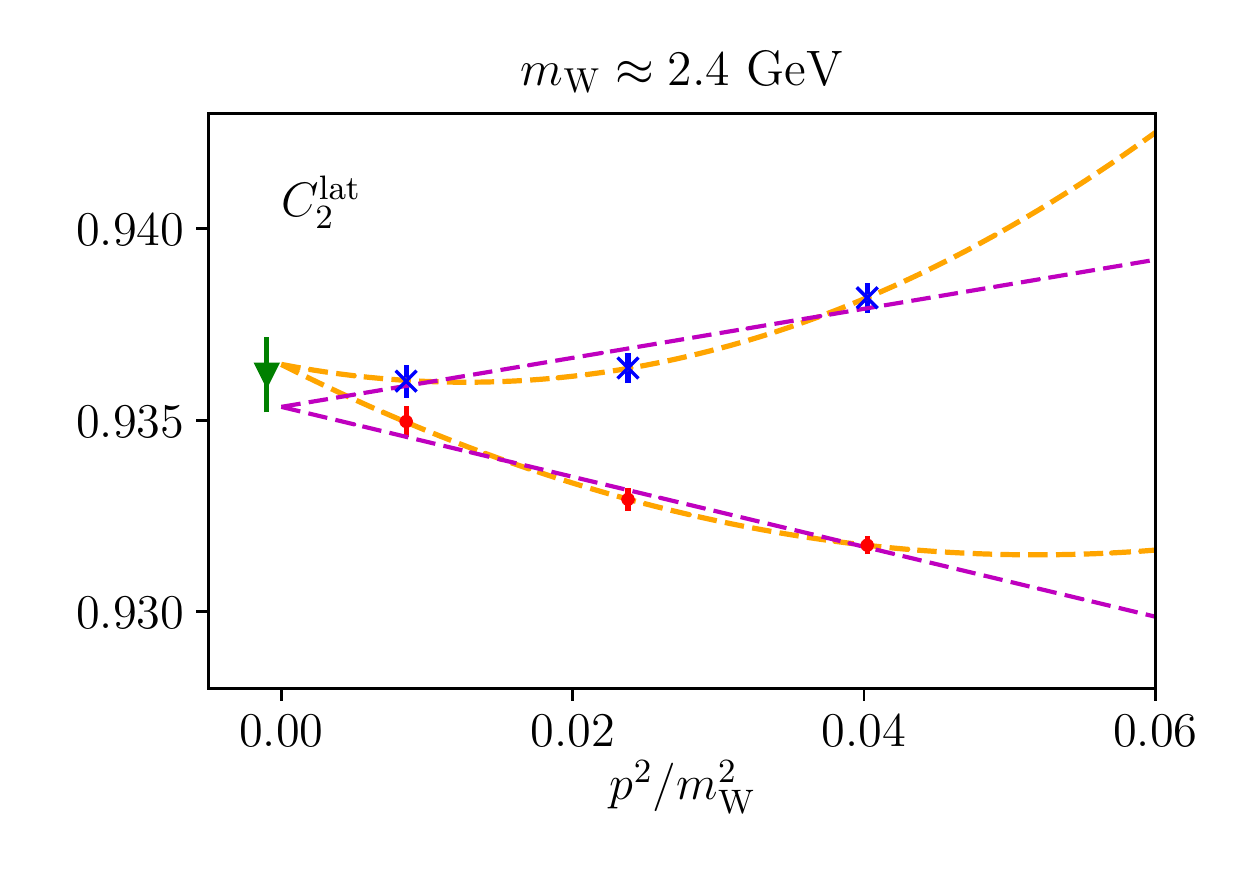}
\includegraphics[width=.49\textwidth]{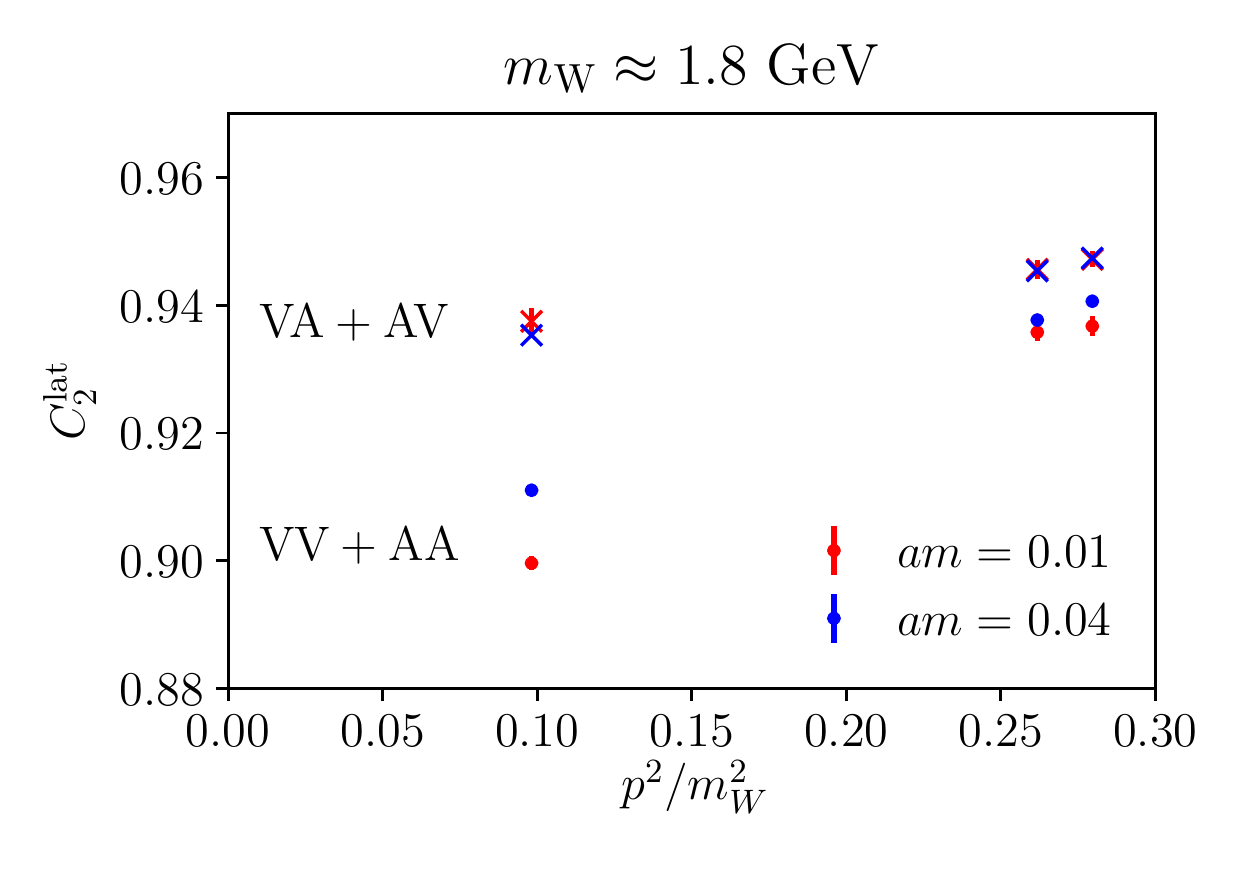}
\caption{
{\it Left}: Bare lattice Wilson coefficient $C_2$
measured on the 32I ensemble with twisted boundary conditions.
Blue crosses correspond to the exceptional kinematics and red dots to
the non-exceptional case. We show combined linear and quadratic extrapolations
in $p^2/\mw^2$ with a common constant term. For the final result, represented
by a green diamond, we use a cubic fit whose larger statistical error covers the 
systematic uncertainty of the extrapolations. The data points have been obtained from
parity odd $\gamma$ projectors.
{\it Right}: $p^2$ dependence of $C_2^\lat$ for two different quark masses and parities of 
$\gamma$ projectors, labeled by the corresponding Dirac structure. A significant pole behavior 
is observed in the parity even case, signaling the contamination from non-perturbative physics.
$\mathcal{CPS}$ symmetry arguments~\cite{Bernard:1985wf,Donini:1999sf} 
also favor the parity odd case, where
mixing with wrong chiralities is suppressed.
Replacing $\gamma$ with $\slashed q$ projectors leads to the same considerations.
}
\label{fig:c2lat}
\end{figure}

The matching procedure with dimension 6 operators becomes exact only in the limit 
$p^2/\mw^2 \to 0$. Therefore in order to control this limit we have computed the amplitudes
$M$ and $W$ for several values of $p$ with exceptional and non-exceptional kinematics.
In the left panel of Figure \ref{fig:c2lat} we present our results for the quantity $C_2^\lat$ 
on the finer ensemble 32I.

For a fixed set of projectors (odd/even and $\gamma$/$\slashed q$)
we have verified that independent fits to the data with exceptional and non-exceptional kinematics
give compatible results in the limit $p^2/\mw^2 \to 0$ within less than one standard deviation.
Therefore we have decided to adopt combined fits with a common constant term for our final
extraction, as reported in the left panel of Figure \ref{fig:c2lat}.
With the present calculation we have not been able to separate QCD and weak
scales enough, as $\Lambda_\qcd/ \mw \approx 0.15$, forcing us 
to push the calculation at very small momenta of $O(\Lambda_\qcd)$, 
where non-perturbative effects may produce sizable contributions.
In these proceedings we present results for the volume and quark mass dependence, 
which provide a qualitative understanding of these systematic errors. 
We refer the reader to Ref.~\cite{wcoeff}
for a detailed and complete study of $O(\Lambda_\qcd)$ effects.

\subsection{Finite volume effects}

\begin{figure}[ht]
\includegraphics[width=.49\textwidth]{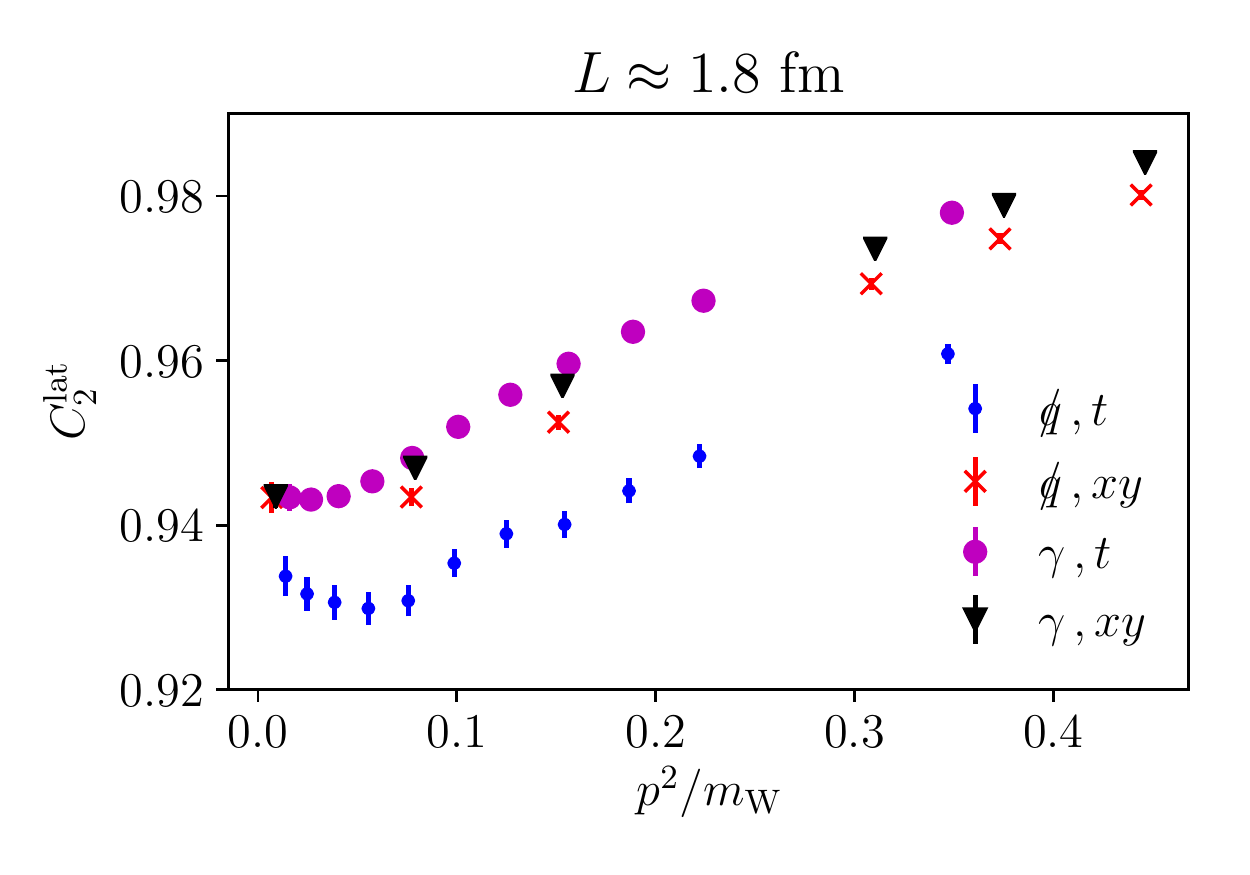}
\includegraphics[width=.49\textwidth]{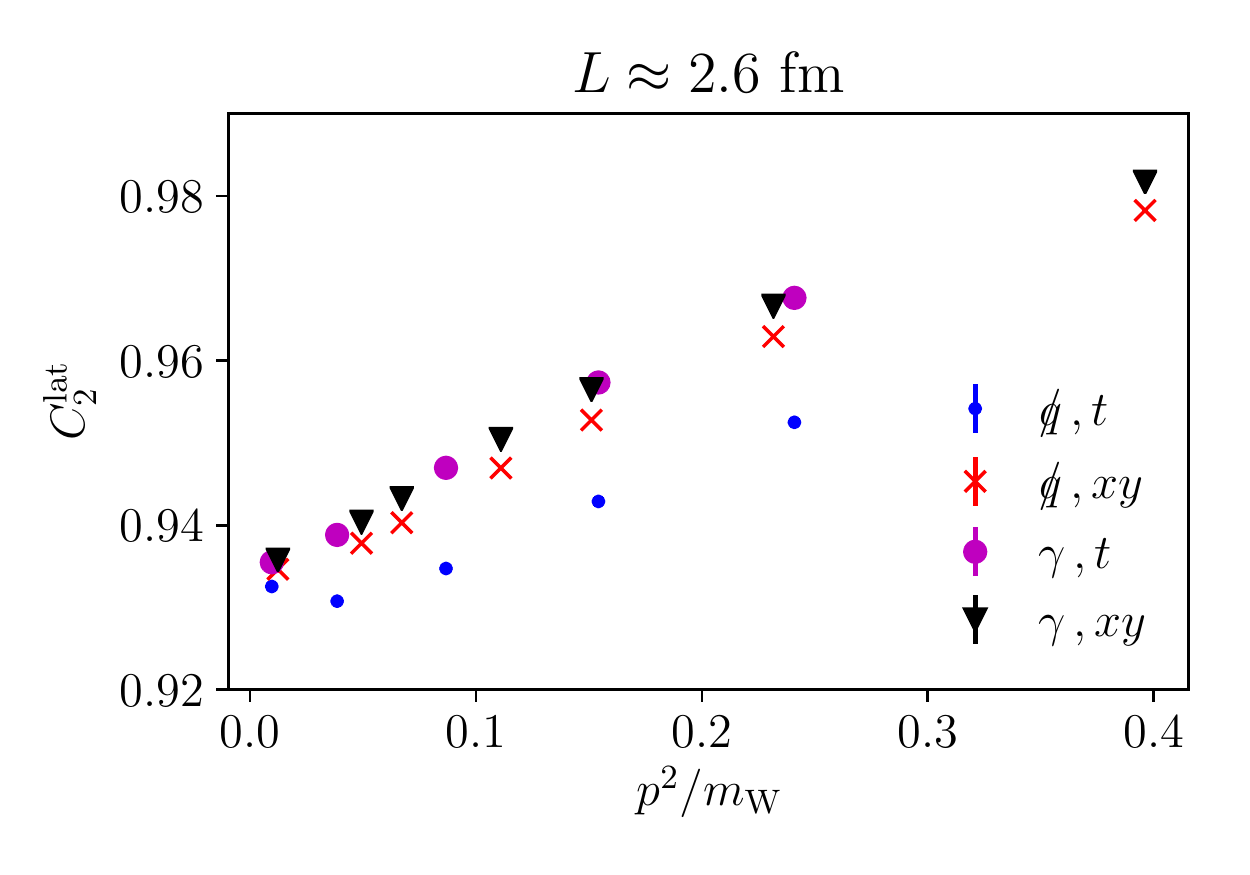}
\caption{Wilson coefficient $C_2^\lat$ measured on the 16I and 24I ensembles
with the same $W$ boson mass of approximately 1.8 $\GeV$. 
To exploit the different extents of the time and spatial directions, 
we have computed the quark propagators with momenta oriented along
the $xy$ directions and $t$ alone. Results for (parity odd) $\gamma$ and $\slashed q$
projectors are reported as well.
In the left panel the several measurements on the 16I do not agree with each other
in the limit of small momenta, whereas the 24I shows a better
agreement in the same limit.}
\label{fig:c2fv}
\end{figure}

To examine possible finite volume errors, we have repeated the calculation
of the Wilson coefficients on the 16I and 24I ensembles, differing only in the
volumes with $L\approx 1.8 ~\fm$ and $2.6~\fm$. Our results are plotted in the
two panels of Figure \ref{fig:c2fv}, where we show measurements with 
exceptional kinematics only, from quark propagators with momenta injected along the
$xy$ spatial directions and time ($t$). 
These choices explore the fact that each lattice 
has a temporal extent longer than the spatial one.
In fact, for the 16I ensemble a noticeable difference in the limit of small $p^2/\mw^2$ 
is visible with $\slashed q$ projectors between the two orientations of momenta; 
moreover when momentum is injected along the time direction the data points nicely 
overlap on the corresponding measurements on the 24I lattice,
for which all different estimates agree with each other in the limit of small momenta.
We interpret this behavior as the absence of finite volume errors: their presence
spoils the convergence in the $p^2 \to 0$ limit for fixed projectors and kinematics.

\subsection{Finite quark mass errors}

The last infrared scale that we analyze in these proceedings is the quark mass. 
Ideally we would like to perform our calculation in a theory with mass-less quarks, 
but in practice 
all propagators have been computed at the strange quark mass of the corresponding
ensemble. Hence, on the 24I we have studied the quark mass dependence by repeating the 
measurements with exceptional kinematics with $am=0.01$.
The results are reported in the
right panel of Figure \ref{fig:c2lat}.
In this case the parity of the projectors plays a very important role.
In the plot, where we show data only for allowed fourier modes,
we observe a good agreement for parity odd projectors between the two masses studied and
a rather strong behavior, approximately of the form $1/p^2$, for the opposite parity.
In particular we note that increasing the quark mass suppresses the pole behavior, 
which we interpret as a non-perturbative effect due to the contamination of pseudo-scalar
mesons to the amplitudes.
For the central values of the Wilson coefficients we use the results from the parity odd
sector, which in general is expected to show less contamination due to the 
suppression of the mixing with the wrong chiralities from $\mathcal{CPS}$ 
symmetry~\cite{Bernard:1985wf,Donini:1999sf}.
The difference between the two parities can be used as an estimate for the systematic
uncertainties of the calculation, which turns out to be rather large ($\approx 2-3 \% $)
compared to the excellent statistical precision, which is below a percent.
Residual chiral symmetry breaking effects are controlled by the separation of the two
walls in the fifth dimension: for a few points we recomputed $C_1^\lat$ and $C_2^\lat$ 
with $L_5=32$ and we observed an excellent agreement with the
original calculation at $L_5=16$ for all projectors.

\subsection{Renormalization and continuum limit}

For a comparison against known results in the continuum limit 
the appropriate renormalization factors have to be computed.
First, we fix four values of the $W$ boson mass in physical units
and we use the ratio of lattice spacings to tune them between
the 24I and 32I ensembles.
Second, to reproduce the initial conditions of the Wilson coefficients
we measure the $Z$ factors at a renormalization
scale equivalent to the four values of $\mw$. Twisted boundary
conditions allow us to reach these points in momentum space without problems
and in the evaluation of the $Z$ factors we adopt the RI/SMOM conditions
defined in Refs.~\cite{Sturm:2009kb,Lehner:2011fz}.
We also take the mass-less limit by repeating the calculation at two different quark masses
and we perform a linear extrapolation in $m$.

The continuum limit is one of the important systematic effects to be addressed here. 
If the $W$ boson is too heavy compared to the cutoff, 
large discretization effects can appear in the Wilson coefficients, 
which essentially capture the modes around and above $\mw$.
The four values of the $W$ mass used here are
$O(1)$ in lattice units.
For $C_2^\ri$ a small lattice spacing error is found in the continuum extrapolations, well
below 1\%, while for $C_1^\ri$ it is larger, around 15\%, partly due to the fact that
the central value is much smaller ($C_1$ is 0 at tree-level).

\begin{figure}[ht]
\centering
\includegraphics[width=.5\textwidth]{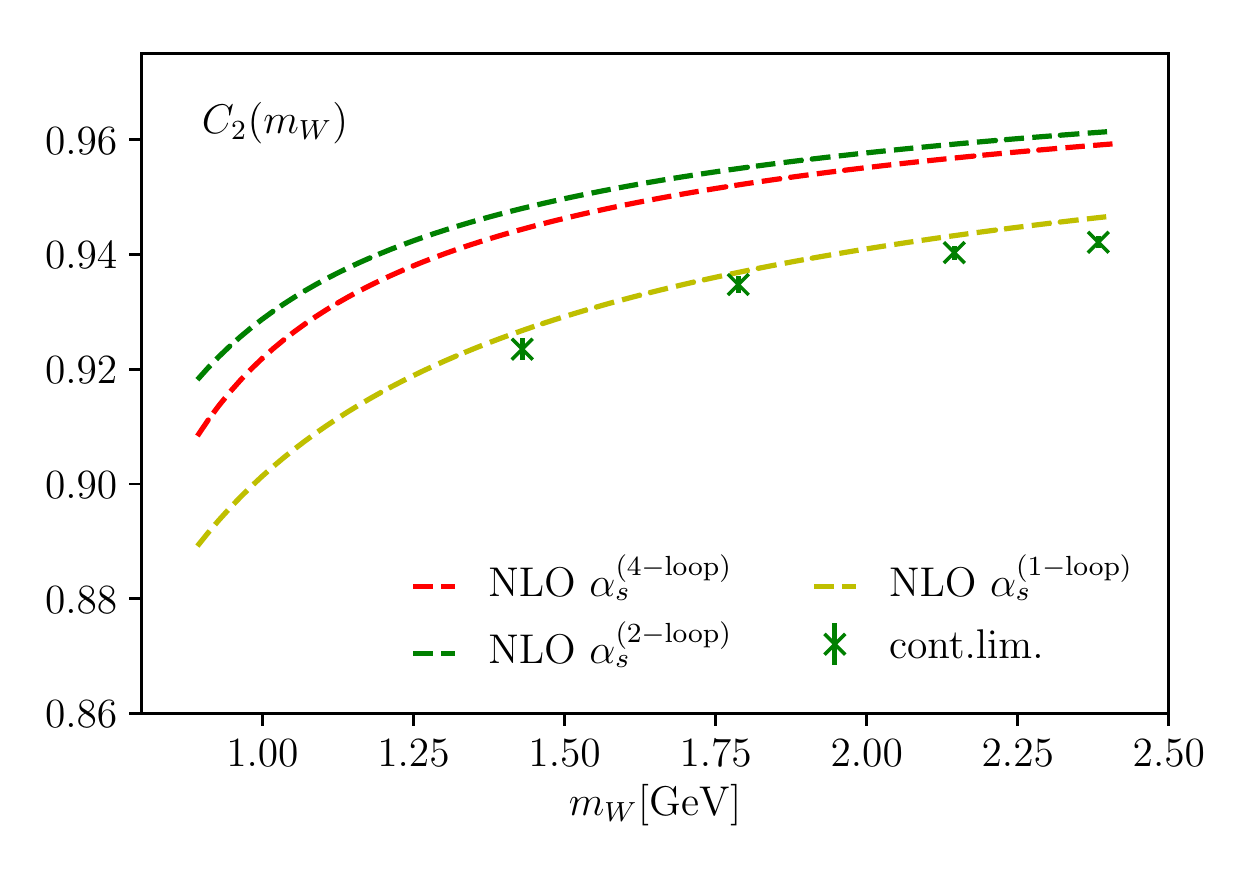}
\caption{The data points are our results in the continuum 
limit for $C_2$ in the RI/SMOM scheme with $\gamma$ projectors.
Only statistical errors are showed. The dashed lines correspond to the known
1-loop prediction in the $\ms$ scheme with three definitions of the strong coupling
constant obtained from 1, 2 and 4-loop $\beta$ functions ($\ms$).}
\label{fig:c2ri}
\end{figure}

Finally we present some preliminary results for the Wilson coefficients in the $\ri$ scheme, 
together with their prediction from perturbation theory in the $\ms$ scheme. Note that the
conversion factor $Z^{\ri \to \ms}$ is currently missing from the calculation, an issue
which we address in Ref.~\cite{wcoeff} for both $\gamma$ and $\slashed q$ intermediate
schemes. 
For a qualitative comparison we present 
in Figure~\ref{fig:c2ri} 
the results from the $\gamma$ scheme where
we expect the conversion matrix $Z^{\ri \to \ms}$ to be very close to the identity~\cite{Lehner:2011fz}.
An agreement with the perturbative predictions is found on a few percent level.
We only display the statistical error bars
without any estimate of systematic uncertainties
or additional non-perturbative errors due to $O(\Lambda_\qcd)$ effects.
Recall that the difference between parity odd and even determinations, which is 
a good representative of such an error, is around 0.03 in absolute units.

The naive perturbative expansion of the Wilson coefficients
in powers of $\alpha_s$ can be used, in principle, 
to determine the various coefficients 
of the first $n$-loop terms.
In the situation where all systematic uncertainties have been addressed properly, 
fitting the Wilson coefficients reported in Figure \ref{fig:c2ri} 
with a functional form like
$C_2^\ri = 1 + k_0 \alpha_s + k_1 \alpha_s^2 + \dots$
(and similarly for $C_1$ with tree-level value equal to 0)
would allow us to
predict them in the physical scenario, by running $\alpha_s$ to the physical $W$ pole,
and to provide a concrete bound on the error 
of the initial conditions. Besides the difficulties
mentioned above, a second caveat to take into account in such fits 
would be the flavor-dependence of the
coefficients starting from two loops, which would require at least two simulations differing
 in the number of flavors in the sea. More details on how the current calculation 
 can be extended to the physical scenario is given in Ref.~\cite{wcoeff}.

\section{Conclusions}

In these proceedings we have presented a method to compute the Wilson coefficients
for the weak effective hamiltonian from lattice simulations. 
We have adopted a RI scheme and explored both the exceptional and non-exceptional
kinematics in the extraction of the bare Wilson coefficients.
By studying their dependence on the lattice size, on the quark mass and on
the external states through different projectors, we have observed large
non-perturbative effects that have to be included in the final systematic 
uncertainties.

We have demonstrated how our numerical strategy leads to results
with small statistical errors.
Already at this preliminary stage, 
without properly accounting for the differences in the renormalization
schemes, we observe a relative good agreement with the known perturbative predictions.
In a second publication we further advance this 
calculation by properly addressing all the systematic errors, 
with a more detailed and comprehensive study~\cite{wcoeff}.

\bibliography{PoS2017.bib}

\begin{thebibliography}{18}

\bibitem{buras95}
G.~Buchalla, A.J. Buras, M.E. Lautenbacher, Rev. Mod. Phys. \textbf{68}, 1125
  (1996), \texttt{hep-ph/9512380}

\bibitem{Bai:2015nea}
Z.~Bai et~al. (RBC, UKQCD), Phys. Rev. Lett. \textbf{115}, 212001 (2015),
  \texttt{1505.07863}

\bibitem{Blum:2012uk}
T.~Blum et~al., Phys. Rev. \textbf{D86}, 074513 (2012), \texttt{1206.5142}

\bibitem{Blum:2011ng}
T.~Blum et~al., Phys. Rev. Lett. \textbf{108}, 141601 (2012),
  \texttt{1111.1699}

\bibitem{CKelly}
C.~Kelly, EPJ Web Conf. \textbf{LATTICE2017} (2017)

\bibitem{Gorbahn:2004my}
M.~Gorbahn, U.~Haisch, Nucl. Phys. \textbf{B713}, 291 (2005),
  \texttt{hep-ph/0411071}

\bibitem{Dawson:1997ic}
C.~Dawson, G.~Martinelli, G.C. Rossi, C.T. Sachrajda, S.R. Sharpe, M.~Talevi,
  M.~Testa, Nucl. Phys. \textbf{B514}, 313 (1998), \texttt{hep-lat/9707009}

\bibitem{Martinelli:1994ty}
G.~Martinelli, C.~Pittori, C.T. Sachrajda, M.~Testa, A.~Vladikas, Nucl. Phys.
  \textbf{B445}, 81 (1995), \texttt{hep-lat/9411010}

\bibitem{Aoki:2007xm}
Y.~Aoki et~al., Phys. Rev. \textbf{D78}, 054510 (2008), \texttt{0712.1061}

\bibitem{Sturm:2009kb}
C.~Sturm, Y.~Aoki, N.H. Christ, T.~Izubuchi, C.T.C. Sachrajda, A.~Soni, Phys.
  Rev. \textbf{D80}, 014501 (2009), \texttt{0901.2599}

\bibitem{Lehner:2011fz}
C.~Lehner, C.~Sturm, Phys. Rev. \textbf{D84}, 014001 (2011), \texttt{1104.4948}

\bibitem{deDivitiis:2004kq}
G.M. de~Divitiis, R.~Petronzio, N.~Tantalo, Phys. Lett. \textbf{B595}, 408
  (2004), \texttt{hep-lat/0405002}

\bibitem{Arthur:2010ht}
R.~Arthur, P.A. Boyle (RBC, UKQCD), Phys. Rev. \textbf{D83}, 114511 (2011),
  \texttt{1006.0422}

\bibitem{Aoki:2010dy}
Y.~Aoki et~al. (RBC, UKQCD), Phys. Rev. \textbf{D83}, 074508 (2011),
  \texttt{1011.0892}

\bibitem{Allton:2007hx}
C.~Allton et~al. (RBC, UKQCD), Phys. Rev. \textbf{D76}, 014504 (2007),
  \texttt{hep-lat/0701013}

\bibitem{Bernard:1985wf}
C.W. Bernard, T.~Draper, A.~Soni, H.D. Politzer, M.B. Wise, Phys. Rev.
  \textbf{D32}, 2343 (1985)

\bibitem{Donini:1999sf}
A.~Donini, V.~Gimenez, G.~Martinelli, M.~Talevi, A.~Vladikas, Eur. Phys. J.
  \textbf{C10}, 121 (1999), \texttt{hep-lat/9902030}

\bibitem{wcoeff}
M.~Bruno, C.~Lehner, A.~Soni (2017), \texttt{1711.05768}

\end{thebibliography}

\end{document}